\documentclass[10pt,letterpaper,aps,pra,floatfix,longbibliography,reprint,amsmath,amsfonts,amssymb,superscriptaddress]{revtex4-1}
\usepackage{braket}
\usepackage{empheq}
\usepackage[normalem]{ulem}

\usepackage[pdfstartview=FitH]{hyperref}
\hypersetup{
    colorlinks=true,       
    linkcolor=cyan,          
    citecolor=magenta,        
    filecolor=magenta,      
    urlcolor=cyan,           
    runcolor=cyan
}

\def\la{\langle}
\def\ra{\rangle}

\begin{document}
\title{Strongly measuring qubit quasiprobabilities behind out-of-time-ordered correlators}

\author{Razieh Mohseninia}
\affiliation{Institute for Quantum Studies, Chapman University, Orange, CA 92866, USA}
\affiliation{Departments of Electrical Engineering, University of Southern California, Los Angeles, California 90089, USA}

\author{Jos\'e Ra\'ul Gonz\'alez Alonso}
\affiliation{Schmid College of Science and Technology, Chapman University, Orange, CA 92866, USA}

\author{Justin Dressel}
\affiliation{Institute for Quantum Studies, Chapman University, Orange, CA 92866, USA}
\affiliation{Schmid College of Science and Technology, Chapman University, Orange, CA 92866, USA}

\date{\today}

\begin{abstract}
Out-of-time-ordered correlators (OTOCs) have been proposed as a tool to witness quantum information scrambling in many-body system dynamics. These correlators can be understood as averages over nonclassical multi-time quasi-probability distributions (QPDs). These QPDs have more information, and their nonclassical features witness quantum information scrambling in a more nuanced way. However, their high dimensionality and nonclassicality make QPDs challenging to measure experimentally. We focus on the topical case of a many-qubit system and show how to obtain such a QPD in the laboratory using circuits with three and four sequential measurements. Averaging distinct values over the same measured distribution reveals either the OTOC or parameters of its QPD. Stronger measurements minimize experimental resources despite increased dynamical disturbance.
\end{abstract}

\maketitle

\section{Introduction}\label{sec:introduction}
The out-of-time-ordered correlator (OTOC) has attracted considerable recent attention in high energy physics \cite{Shenker_BlackHolesButterfly_2014,Shenker_MultipleShocks_2014,Shenker_StringyEffectsScrambling_2015,Roberts_LocalizedShocks_2015,Roberts_DiagnosingChaosUsing_2015,Maldacena_BoundChaos_2016,Stanford_ManybodyChaosWeak_2016,Maldacena_RemarksSachdevYeKitaevModel_2016,Blake_UniversalChargeDiffusion_2016,Blake_UniversalDiffusionIncoherent_2016,Roberts_LiebRobinsonBoundButterfly_2016,Hosur_ChaosQuantumChannels_2016,Lucas_ChargeDiffusionButterfly_2016,Chen_UniversalLogarithmicScrambling_2016,Gu_LocalCriticalityDiffusion_2017} and condensed matter physics \cite{Aleiner_MicroscopicModelQuantum_2016,Banerjee_SolvableModelDynamical_2017,Huang_OutoftimeorderedCorrelatorsManybody_2017,Swingle_SlowScramblingDisordered_2017,Fan_OutoftimeorderCorrelationManybody_2017,Patel_QuantumButterflyEffect_2017,Chowdhury_OnsetManybodyChaos_2017,He_CharacterizingManybodyLocalization_2017,Patel_QuantumChaosCritical_2017,Kukuljan_WeakQuantumChaos_2017,Lin_OutoftimeorderedCorrelatorsQuantum_2018}. It helps characterize quantum information scrambling due to the spread of entanglement, and has found utility in applications ranging from black hole thermalization to quantum chaos. Alongside the theoretical effort, there has been increasing interest in finding experimental methods to measure such a quantity in modern quantum simulators (e.g., \cite{Swingle_MeasuringScramblingQuantum_2016,Zhu_MeasurementManybodyChaos_2016,Danshita_CreatingProbingSachdevYeKitaev_2017,Li_MeasuringOutofTimeOrderCorrelators_2017,Garttner_MeasuringOutoftimeorderCorrelations_2017}). These controllable quantum systems may be used to simulate and measure exotic dynamics that are otherwise out of experimental reach, such as quantum state teleportation through a traversable wormhole \cite{Yoshida_DisentanglingScramblingDecoherence_2019,Landsman_VerifiedQuantumInformation_2019}.

Expanding upon the idea of the OTOC, we recently introduced a more refined and robust information-scrambling witness by decomposing the OTOC into its extended (coarse-grained) Kirkwood-Dirac \cite{Kirkwood_QuantumStatisticsAlmost_1933,Dirac_AnalogyClassicalQuantum_1945,Lundeen_DirectMeasurementQuantum_2011,Mirhosseini_CompressiveDirectMeasurement_2014,Lundeen_ProcedureDirectMeasurement_2012,Bamber_ObservingDiracClassical_2014,Dressel_WeakValuesInterference_2015} quasiprobability distribution (QPD) \cite{YungerHalpern_JarzynskilikeEqualityOutoftimeordered_2017,YungerHalpern_QuasiprobabilityOutofTimeOrderedCorrelator_2018}. This QPD has since found utility in entropic uncertainty relations for scrambling \cite{YungerHalpern_EntropicUncertaintyRelations_2018}, and is closely related to a witness for quantum advantage in postselected metrology \cite{Arvidsson-Shukur_ContextualityProvidesQuantum_2019}. The OTOC signals interesting scrambling behavior when it decays to a persistently small value; to produce this decay, its associated QPD must exhibit negative or non-real values, despite satisfying all other properties of a probability distribution. The OTOC is an average over this QPD, so it has less information than the full QPD about the probed system dynamics. Moreover, while the OTOC can also decay due to decoherence in a manner that seems qualitatively similar to the decay from information scrambling, the nonclassical features of the corresponding QPD can only diminish with decoherence. As such, the QPD robustly distinguishes such decoherence from scrambling \cite{GonzalezAlonso_OutofTimeOrderedCorrelatorQuasiprobabilitiesRobustly_2019}, making it an attractive candidate for experimental use.

The apparent problem with the QPD is that it is a 4-argument distribution, and thus seems to require the experimental measurement of many more parameters than the OTOC. Indeed, for a qubit OTOC there are 2 real parameters to measure, but its corresponding QPD ostensibly has $2\times2^4 = 32$ real parameters in the distribution. Without a practical method of determining all the parameters composing the QPD, its advantages compared to an OTOC are reduced.

In this paper we show that a qubit QPD can be measured using the same sequential measurement circuit used to determine the OTOC itself, which demonstrates that it is no more difficult to measure in spite of its high-dimensionality. We accomplish this feat through two simplification steps: First, we show that the 32 real parameters of the QPD are redundant and can be reduced to 8 independent correlators. Second, we generalize the method that we introduced in Ref.~\cite{Dressel_StrengtheningWeakMeasurements_2018} for measuring qubit OTOCs using 2 circuits of sequential measurements. We show that the same circuits also yield all 8 correlators that determine the QPD. Moreover, the statistical error is minimal when all but the first measurement are projective, with the first only slightly weakened.

This paper is organized as follows. In Section~\ref{sec:otoc} we review the OTOC and its associated QPD. In Section~\ref{sec:qubits} we review sequential qubit measurements and the key results of Ref.~\cite{Dressel_StrengtheningWeakMeasurements_2018}. In Section~\ref{sec:qpdmeasure} we detail how to measure the QPD efficiently. In Section~\ref{sec:optimization} we optimize the measurement strengths to minimize statistical error. We conclude in Section~\ref{sec:conclusion}.

\section{OTOCs and their QPDs}\label{sec:otoc}
We consider the important case of a lattice of locally interacting qubits, such as those used in modern quantum computing hardware. When such a multi-qubit system evolves with a Hamiltonian $H$, the dynamics can cause initially localized information to spread through the lattice. More precisely, an initially localized single-qubit operator $A$ will typically evolve to have support over multiple lattice sites in the Heisenberg picture, $A(t) = U^\dagger(t) A U(t)$, with $i\hbar\partial_tU(t)= HU(t)$ and $U(0)=I$. Integrable Hamiltonians cause periodic evolution that will relocalize such an operator at a future recurrence time. However, non-integrable Hamiltonians can have an exponentially longer recurrence time \cite{Bocchieri_QuantumRecurrenceTheorem_1957,Hosur_ChaosQuantumChannels_2016,CamposVenuti_RecurrenceTimeQuantum_2015} that persistently scrambles the information of the initially local operators to cover the lattice. An OTOC and its QPD can witness such information-scrambling behavior \cite{GonzalezAlonso_OutofTimeOrderedCorrelatorQuasiprobabilitiesRobustly_2019}.

We assume in this paper that local qubit operators $A$ and $B$ at distinct lattice sites square to the identity $A^2 = B^2 = I$ and initially commute $[A,B] = 0$. At later times $t$, however, $B(t)$ can evolve to overlap the initial support of $A$. We can detect such emergent overlap by averaging the positive Hermitian-square of their commutator after evolving only $B$,
\begin{align}
C(t) := \big \la [A,B(t)]^\dagger [A,B(t)]  \big \ra = 2[1-\text{Re}F(t)] \geq 0.
\end{align}
Since $A^2 = B(t)^2 = I$ for any $t$, $C(t)$ is determined by
\begin{align} \label{eq:otoc}
F(t):= \la B(t) A B(t) A \ra,
\end{align}
which is an OTOC that satisfies $F(0) = 1$ and $\text{Re} F(t) \leq 1$. For a non-integrable Hamiltonian, persistent noncommutativity of $A$ and $B(t)$, i.e., $C(t)>0$, causes $\text{Re}F(t)$ to drop to a small value for an extended duration \cite{GonzalezAlonso_OutofTimeOrderedCorrelatorQuasiprobabilitiesRobustly_2019}.

The noncommutativity of $A$ and $B(t)$ also precludes the existence of a classical joint probability distribution over their eigenvalues, so prevents the OTOC from being understood as a simple eigenvalue average. Specifically, if we decompose $A$ and $B$ into their eigenprojection operators $\Pi_a^A$ and $\Pi_b^{B(t)}$, $A= \sum_{a=0,1} (-1)^a\, \Pi_a^A$ and $B(t)=\sum_{b=0,1} (-1)^b\, \Pi_b^{B(t)}$, then the OTOC becomes an eigenvalue average
\begin{align}
F(t) = \!\!\!\!\!\!\sum_{b',a',b,a=0,1} \!\!\!\!\!\! (-1)^{b'+a'+b+a}\,\, \tilde{p}_t(b',a',b,a)
\end{align}
over an extended Kirkwood-Dirac QPD \cite{YungerHalpern_JarzynskilikeEqualityOutoftimeordered_2017,YungerHalpern_QuasiprobabilityOutofTimeOrderedCorrelator_2018}
\begin{align}\label{eq:qpd}
\tilde{p}_t(b',a',b,a) := \la \Pi^{B(t)}_{b'} \Pi^A_{a'} \Pi^{B(t)}_{b} \Pi^A_{a} \ra.
\end{align}
The QPD $\tilde{p}_t$ is normalized, $\sum\tilde{p}_t = 1$, and reduces to a classical probability distribution when $A$ and $B(t)$ commute, but can take imaginary and negative values when $A$ and $B(t)$ do not commute. Thus, the interesting behavior of the OTOC $F(t)$ directly corresponds to when the aggregated nonclassicality of the QPD, $N(t) := \sum |\tilde{p}_t| - 1 \geq 0$, becomes nonzero \cite{GonzalezAlonso_OutofTimeOrderedCorrelatorQuasiprobabilitiesRobustly_2019}. This nonclassicality is a witness of information scrambling that is more robust to experimental imperfections than the OTOC itself \cite{GonzalezAlonso_OutofTimeOrderedCorrelatorQuasiprobabilitiesRobustly_2019}.

\section{Sequential qubit measurements}\label{sec:qubits}
We will measure the OTOC and its QPD with sequences of informative and non-informative ancilla-based qubit measurements. Our analysis extends that of Ref.~\cite{Dressel_StrengtheningWeakMeasurements_2018}, which provides explicit implementation circuits and detailed derivations in its appendix.

An informative measurement of a qubit observable $A$ correlates the measured basis of an ancilla qubit with the eigenbasis of $A$. Measuring a result $a=0,1$ on the ancilla then causes (partial) collapse backaction in the basis of $A$. Such a partial collapse modifies the state $\rho \mapsto M_{\phi,a}^{(A)}\rho M_{\phi,a}^{\dagger(A)}$ according to the Kraus operators \cite{Dressel_StrengtheningWeakMeasurements_2018}
\begin{align} \label{eq:infom}
M_{\phi,a}^{(A)} := \frac{1}{\sqrt{2}} [\cos\frac{\phi}{2}I+(-1)^a \sin\frac{\phi}{2} A].
\end{align}
The parameter $\phi\in(0,\pi/2]$ is an angle that sets the measurement strength \cite{Dressel_StrengtheningWeakMeasurements_2018}, with $\phi=\pi/2$ corresponding to a projective measurement of the eigenbasis of $A$, and $\phi\to 0$ corresponding to the weak measurement limit that leaves $\rho$ nearly unperturbed. For any $\phi$, averaging the ancilla-outcome probabilities $P_\phi^A(a)= \text{tr} (M_{\phi,a}^ {(A)} \rho M_{\phi,a}^ {(A)\dagger})$ with the generalized eigenvalues \cite{Dressel_ContextualValuesObservables_2010,Dressel_ContextualvalueApproachGeneralized_2012,Dressel_QuantumInstrumentsFoundation_2013} $\alpha_{\phi,a} = (-1)^a/\sin\phi$ recovers the expectation value $\la A \ra = \sum_{a=0,1}\alpha_{\phi,a}P_\phi^A(a)$.

A noninformative measurement causes phase backaction by entangling the eigenbasis of $A$ with a mutually unbiased basis of the ancilla. Measuring the ancilla then gives no information about $A$, but does produce a measurement-controlled unitary effect  generated by $A$ on the initial state $\rho \mapsto N_{\phi,a}^{(A)}\rho N_{\phi,a}^{\dagger(A)}$, according to the Kraus operators
\begin{align} \label{eq:noninfom}
N_{\phi,a}^{(A)} := \frac{1}{\sqrt{2}} [\cos\frac{\phi}{2}I-i(-1)^a \sin\frac{\phi}{2} A].
\end{align}
As before, the angle $\phi\in(0,\pi/2]$ indicates the measurement strength, ranging from weak perturbations with $\phi \to 0$ to maximally distinct rotations with $\phi = \pi/2$.

Performing a sequence of $n$ informative measurements of observables $A_1, A_2, \ldots, A_n$, implemented by separate ancillas, produces a joint probability distribution
\begin{align}
&P^{A_1, \ldots, A_n}_{\phi_1, \ldots, \phi_n}(a_1,\ldots,a_n) := \\
&\qquad\quad \text{tr}\left(M_{\phi_n,a_n}^{(A_n)}\cdots M_{\phi_1,a_1}^{(A_1)}\rho M_{\phi_1,a_1}^{\dagger(A_1)}\cdots M_{\phi_n,a_n}^{\dagger(A_n)}\right), \nonumber
\end{align}
where $a_i=0,1$ is the outcome of the $i$\textsuperscript{th} measurement. As shown in Ref.~\cite{Dressel_StrengtheningWeakMeasurements_2018}, averaging this joint distribution with the generalized eigenvalues $\alpha_{\phi_i,a_i} = (-1)^{a_i}/\sin\phi_i$ exactly produces a correlation function involving nested anticommutators of $A_1, A_2, \ldots, A_n$:
\begin{align}
 \label{eq:nested}\nonumber
\mathcal{C}^{A_1,\ldots,A_n} &:= \!\!\sum_{a_1, \ldots, a_n} \!\! \alpha_{\phi_1,a_1}\cdots\alpha_{\phi_n,a_n}\,P^{A_1, \ldots,A_n}_{\phi_1, \ldots,\phi_n}(a_1,\ldots,a_n) \nonumber \\
&= \left\la \frac{\{...\{\{A_n,A_{n-1}\},A_{n-2}\}...,A_1\}}{2^{n-1}} \right\ra
\end{align}
for all strength angles $0 < \phi_i \leq \pi/2$.

Replacing only the first informative measurement $M_{\phi_1,a_1}^{(A_1)}$ with a noninformative measurement $N_{\phi_1,\tilde{a}_1}^{(A_1)}$ in a separate circuit produces a modified joint distribution
\begin{align}
&\tilde{P}_{\phi_1,\ldots,\phi_n}^{A_1,\ldots,A_n}(\tilde{a}_1,\ldots,a_n) := \nonumber \\
&\qquad \text{tr}\left( M_{\phi_n,a_n}^{(A_n)}\cdots N_{\phi_1,\tilde{a}_1}^{(A_1)}\,\rho\,N_{\phi_1,\tilde{a}_1}^{\dagger(A_1)}\cdots M_{\phi_n,a_n}^{\dagger(A_n)}\right),
\end{align}
where the notations $\tilde{P}$ and $\tilde{a}_1$ are used as a reminder of the noninformative nature of the first measurement. Averaging in the same way as in Eq.~\eqref{eq:nested} exchanges the outermost anti-commutator with a commutator \cite{Dressel_StrengtheningWeakMeasurements_2018}
\begin{align}
 \label{eq:nested2}\nonumber
\tilde{\mathcal{C}}^{A_1,\ldots,A_n} &:= \!\!\sum_{\tilde{a}_1, \ldots, a_n} \!\! \alpha_{\phi_1,\tilde{a}_1}\cdots\alpha_{\phi_n,a_n}\,\tilde{P}^{A_1, \ldots,A_n}_{\phi_1, \ldots,\phi_n}(\tilde{a}_1,\ldots,a_n) \nonumber \\
&= \left\la \frac{[...\{\{A_n,A_{n-1}\},A_{n-2}\}...,A_1]}{2^{n-1}i} \right\ra.
\end{align}

In Ref.~\cite{Dressel_StrengtheningWeakMeasurements_2018} we showed that the OTOC $F(t)$ is completely determined by four-measurement correlators $\mathcal{C}^{AB(t)AB(t)}$ and $\tilde{\mathcal{C}}^{AB(t)AB(t)}$. We will now analyze sequences of both informative and noninformative measurements more carefully to improve upon this result and obtain all 8 correlators needed to construct the QPD $\tilde{p}_t$.

\begin{figure*}[t]
    \centering
    \includegraphics{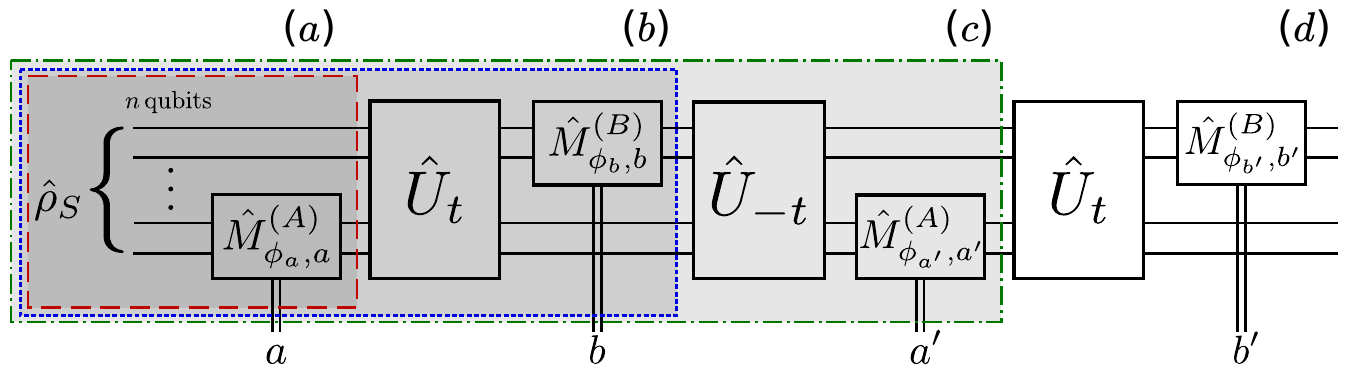}
    \caption{Sequential measurement circuit. Repeated circuit realizations yield the joint distribution $P^{A,B,A,B}_{\phi_a,\phi_b,\phi_{a'},\phi_{b'}}(a,b,a',b')$ of ancilla-qubit outcomes. Averaging this distribution with strategic values (see main text) yields the multi-qubit out-of-time-ordered correlator $F(t) = \left\la B(t) A B(t) A \right\ra$ and 8 correlators that determine its corresponding quasiprobability distribution $\tilde{p}_t$.  (a) Averaging the one-measurement subcircuit (red, dashed) produces $\la A \ra$. (b) Averaging the two-measurement subcircuit (blue, dotted) produces $\text{Re}\la B(t) A\ra$. (c) Averaging the three-measurement subcircuit (green, dot-dashed) produces $\la B(t) \ra$, $\la A B(t) A \ra$, and $\text{Re}\la B(t) A B(t) A\ra$. (d) Averaging the four-measurement circuit produces $\la B(t) A B(t)\ra$. To obtain the final two correlators $\text{Im}\la B(t) A\ra$ and $\text{Im}\la B(t) A B(t) A\ra$ that determine $\tilde{p}_t$, the first informative measurement $\hat{M}^{(A)}_{\phi_a,a}$ should be replaced with a non-informative measurement $\hat{N}^{(A)}_{\phi_a,a}$ (see text for details), and the last measurement may be omitted.}
    \label{fig:measurements}
\end{figure*}

\section{Measuring a QPD}\label{sec:qpdmeasure}
The QPD $\tilde{p}_t$ formally consists of $2^4$ complex numbers, so apparently it requires experimental determination of 32 real parameters. However, we can reduce this complexity to just $8$ real parameters to measure \cite{YungerHalpern_QuasiprobabilityOutofTimeOrderedCorrelator_2018}. Since $A^2 = B^2(t) = I$, we use the identities $\Pi^A_a = [I + (-1)^a A]/2$ and $\Pi^{B(t)}_b = [I + (-1)^b B(t)]/2$ to expand the QPD in Eq.~\eqref{eq:qpd} into
$2^4$ terms that contain only 8 real-valued correlators:
 $\la A \ra$, $\la B(t) \ra$, $\text{Re}\la B(t)A \ra$, $\text{Im}\la B(t)A \ra$, $\la B(t)AB(t) \ra$, $\la AB(t)A \ra$, $\text{Re}\la B(t)AB(t)A \ra$, and $\text{Im}\la B(t)AB(t)A \ra$. Notably, two of these correlators are the real and imaginary parts of the OTOC $F(t)$ itself, emphasizing that the QPD contains more information. Once these 8 independent correlators are determined, the entire QPD may be reconstructed. 

We now consider how to measure each correlator in turn by strategically averaging sequential measurements as in Eqs.~\eqref{eq:nested} and \eqref{eq:nested2}.  Our goal is to measure all needed terms with a minimum amount of experimental resources, including both the number of measurement circuits and the number of realizations of each required to obtain a desired statistical error.

We show that a single circuit with four informative measurements can determine 6 of the 8 correlators. The remaining 2 correlators are determined by a related three-measurement circuit that substitutes the first measurement with a noninformative measurement. To be systematic, we construct the circuit shown in Fig.~\ref{fig:measurements} by adding one measurement at a time.

\subsection{One-measurement sub-circuit}
We start from the smallest sub-circuit in Fig.~\ref{fig:measurements}(a) (red, dashed)  consisting of one informative measurement of $A$. According to Eq.~\eqref{eq:nested}, we obtain $\la A \ra$ by averaging the values
\begin{align}
    \xi^{A}_a &:= \alpha_{\phi_a,a} \equiv \frac{(-1)^a}{\sin\phi_a}
\end{align}
over the distribution $P^A_{\phi_a}(a)$. We show later that the other single-point correlator (i.e., expectation value) $\la B(t) \ra$ can be obtained by the three-measurement sub-circuit in Fig.~\ref{fig:measurements}(c) (green, dot-dashed).

\subsection{Two-measurement sub-circuit}
Adding an informative measurement of $B(t)$ produces the two-measurement sub-circuit in Fig.~\ref{fig:measurements}(b) (blue, dotted). As discussed in Ref~\cite{Dressel_StrengtheningWeakMeasurements_2018}, measuring $B(t)$ requires first evolving the qubit system for a duration $t$, then coupling the eigenspace of $B$ to an ancilla, then backward-evolving for a duration $t$. The backwards evolution may be omitted if it occurs at the end of the subcircuit. According to Eq.~\eqref{eq:nested}, averaging the simple product
\begin{align}
    \xi^{AB}_{a,b} := \alpha_{\phi_a,a}\,\alpha_{\phi_b,b}
\end{align}
over the joint distribution $P^{A,B}_{\phi_a,\phi_b}(a,b)$ produces the correlator $\mathcal{C}^{A,B} = \la \{B(t),A\} \ra/2 = \text{Re}\la B(t)A \ra$. Substituting the first measurement with a non-informative measurement as in Eq.~\eqref{eq:nested2} and averaging the same values $\xi^{AB}_{a,b}$ yields $\tilde{\mathcal{C}}^{A,B} = \la [B(t),A] \ra/2i = \text{Im}\la B(t) A \ra$ instead \cite{Dressel_StrengtheningWeakMeasurements_2018}. For brevity, we omit the time-dependence of $B(t)$ in the remainder of the paper as understood.

To elucidate the structure of this sub-circuit, we compute the measured distribution $P_{\phi_a,\phi_b}^{A,B}(a,b)$. Using Eq.~\eqref{eq:infom} we find
\begin{align}\label{eq:twomeasurement}
&P^{A,B}_{\phi_a,\phi_b}(a,b) = \frac{1}{4} \bigg[1+(-1)^a \sin\phi_a\, \la A \ra \nonumber \\
&\quad\quad + (-1)^b \sin\phi_b\, \left( \cos^2\frac{\phi_a}{2}\, \la B \ra  + \sin^2\frac{\phi_a}{2}\, \la ABA \ra\right) \nonumber \\
&\quad\quad + (-1)^{a+b} \sin\phi_a\, \sin\phi_b \frac{\la \{B,A \} \ra }{2} \bigg].
\end{align}
This form shows that marginalizing over $b=0,1$ cancels the last two lines to recover the result for the one-measurement sub-circuit. However, marginalizing over $a=0,1$ and averaging $b$ with the generalized eigenvalues $\alpha_{\phi_b,b} = (-1)^b/\sin\phi_b$ only cancels the terms with $\la A \ra$ and $\la \{B,A\}\ra$ to leave a linear combination of $\la B \ra$ and $\la A B A \ra$, making it impossible to isolate those two correlators independently. Intuitively, the first measurement of $A$ (partially) collapses the state, which correlates the result of the second measurement with the first.

Note that if we perform a \emph{weak} measurement of the observable $A$ with $\phi_a \approx 0$, then the pre-factor of $\la ABA \ra$ in Eq.~\eqref{eq:twomeasurement} becomes negligible compared to $\la B \ra$ because it is quadratic in $\phi_a$. In this case, the marginalization of Eq.~\eqref{eq:twomeasurement} approximates $P(b)$, from which we can isolate $\la B \ra$. However, weak measurements require more experimental realizations to minimize statistical error, so instead we will directly isolate both $\la B \ra$ and $\la A B A \ra$ after adding one more measurement of $A$.

\subsection{Three-measurement sub-circuit}
Adding an informative measurement of $A$ yields the three-measurement sub-circuit in Fig.~\ref{fig:measurements}(c) (green, dot-dashed). The joint probability distribution of the measured outcomes is then $P^{A,B,A}_{\phi_a,\phi_b,\phi_{a'}}(a,b,a')$. The structure of this distribution is similar to that of Eq.~\eqref{eq:twomeasurement}, but we omit its full form for brevity. This joint distribution will allow us to obtain the correlators $\la B \ra$, $\la ABA \ra$, and $\text{Re}\la BABA \ra$, while the modified distribution $\tilde{P}^{A,B,A}_{\phi_a,\phi_b,\phi_{a'}}(\tilde{a},b,a')$ will produce $\text{Im}\la BABA\ra$.

Following Eq.~\eqref{eq:nested}, averaging $P^{A,B,A}_{\phi_a,\phi_b,\phi_{a'}}(a,b,a')$ with the product $\alpha_{\phi_a,a}\,\alpha_{\phi_b,b}\,\alpha_{\phi_{a'},a'}$ produces the correlator $\mathcal{C}^{A,B,A} = \la \{\{A,B
\},A
\}/4 \ra = \la B+ABA \ra /2$. This result produces a second linear combination of $\la B \ra$ and $\la ABA \ra$, which we can combine with a partial average of Eq.~\eqref{eq:twomeasurement} to isolate both $\la B \ra$ and $\la A B A \ra$ separately. Solving this linear system to obtain $\la B \ra$ yields the effective values
\begin{align}
\label{eq:Bvalue}
 \xi^B_{a,b,a'} &:= \frac{\alpha_{\phi_b,b} - 2\,\alpha_{\phi_a,a}\,\alpha_{\phi_b,b}\,\alpha_{\phi_{a'},a'}\,\sin^2(\phi_a/2)}{\cos\phi_a}
\end{align}
to average over the distribution $P^{A,B,A}_{\phi_a,\phi_b,\phi_{a'}}(a,b,a')$. Similarly, to obtain $\la ABA \ra$ we average the values
\begin{align}
\xi^{ABA}_{a,b,a'} := 2\,\alpha_{\phi_a,a}\,\alpha_{\phi_b,b}\,\alpha_{\phi_{a'},a'} - \xi^B_{a,b,a'}.
\end{align}

We note two important subtleties of this result. First, $B$ may be isolated in the measurement sequence $(A, B, A)$ because the first $A$ measurement algebraically cancels with the final $A$ measurement, which is only possible because $A^2 = I$. Surprisingly, the later measurement allows us to ``undo'' the effect of the earlier measurement. Second, this cancellation is only possible when the first measurement is \emph{not} projective, $\phi_a \neq \pi/2$. Intuitively, a projective measurement would irreversibly collapse the state, preventing information from being retrieved and canceled. However, cancellation is possible with any other measurement strength $0 < \phi_a < \pi/2$.

In addition to $\la B \ra$ and $\la ABA \ra$, we can also obtain the OTOC itself $\text{Re}\la BABA \ra$ from the distribution $P^{A,B,A}_{\phi_a,\phi_b,\phi_{a'}}(a,b,a')$. Much as $P^{A,B}_{\phi_a,\phi_b}(a,b)$ in Eq.~\eqref{eq:twomeasurement} contains $\la ABA \ra$, the OTOC appears in backaction terms. To extract $\text{Re}\la BABA \ra$ directly, we average the values
\begin{align}\label{eq:rebaba}
    \xi^{\text{Re}BABA}_{a,b,a'} := \frac{\alpha_{\phi_a,a}\,\alpha_{\phi_{a'},a'} - \cos^2(\phi_b/2)}{\sin^2(\phi_b/2)}
\end{align}
over the joint distribution $P^{A,B,A}_{\phi_a,\phi_b,\phi_{a'}}(a,b,a')$. This result simplifies the OTOC-measuring protocol in Ref.~\cite{Dressel_StrengtheningWeakMeasurements_2018} by removing the need for a fourth measurement.

To extract the imaginary part of the OTOC $\text{Im}\la BABA\ra$ we replace $M^{(A)}_{\phi_a,a}$ with $N^{(A)}_{\phi_{\tilde{a}},\tilde{a}}$ in Fig.~\ref{fig:measurements}(c) and average the values
\begin{align}\label{eq:imbaba}
    \xi^{\text{Im}BABA}_{\tilde{a},b,a'} := \frac{\alpha_{\phi_{\tilde{a}},\tilde{a}}\,\alpha_{\phi_{a'},a'}}{\sin^2(\phi_b/2)}
\end{align}
over the modified joint distribution $\tilde{P}^{A,B,A}_{\phi_{\tilde{a}},\phi_b,\phi_{a'}}(\tilde{a},b,a')$.

So far we have obtained 7 of the 8 correlators needed to determine the OTOC QPD, with only $\la BAB \ra$ remaining. Unfortunately, the three-measurement circuit is not sufficient for the same reason that $\la B \ra$ could not be obtained from the sequence $(A, B)$ in Eq.~\eqref{eq:twomeasurement}.
That is, after marginalizing $a$ and $b$ then averaging $a'$ we find
\begin{align} \label{eq:margin3}
&\sum_{a, b, a'}  \alpha_{\phi_{a'},a'} P^{A,B,A}_{\phi_a,\phi_b,\phi_{a'}}(a, b,a') =\cos^2\frac{\phi_b}{2} \la A \ra \\ \nonumber
&\quad+ \sin^2\frac{\phi_b}{2}\cos^2\frac{\phi_a}{2} \la BAB \ra +\sin^2\frac{\phi_b}{2}\sin^2\frac{\phi_a}{2} \la ABABA \ra.
\end{align}
The correlator $\la BAB \ra$ appears in a linear combination with both $\la A \ra$ and $\la ABABA \ra$, so can not be isolated unless the first measurement is made weak with $\phi_a \approx 0$.

\subsection{Four-measurement circuit}
Adding one last informative measurement of $A$ produces the full circuit in Fig.~\ref{fig:measurements}(d). The remaining $\la BAB \ra$ correlator can then be isolated. As with the $\la B \ra$ correlator, the effect of the first $A$ measurement is undone by subsequent measurements; however, the cancellation is more complicated and involves measurement backaction terms similarly to the OTOC correlators in the previous section. To extract $\la BAB \ra$, we average the values
\begin{align}
\xi^{BAB}_{a,b,a',b'} := \frac{1}{\cos \phi_a} \bigg[ & - \alpha_{\phi_{a},a} + 2\,\alpha_{\phi_{b},b}\,\alpha_{\phi_{a'},a'}\,\alpha_{\phi_{b'},b'} \\
&- 2\, \alpha_{\phi_{a},a}\, \alpha_{\phi_{b},b}\, \alpha_{\phi_{b'},b'} \frac{\sin^2(\phi_a/2)}{\sin^2 (\phi_{a'}/2)} \nonumber \\
&+ 2\, \alpha_{\phi_{a},a} \frac{\sin^2 (\phi_a/2) \cos^2(\phi_{a'}/2)}{\sin^2 (\phi_{a'}/2)}   \bigg] \nonumber
\end{align}
over the joint distribution $P^{A,B,A,B}_{\phi_a,\phi_b,\phi_{a'},\phi_{b'}}(a,b,a',b')$. As with the correlator $\la B \ra$, needed cancellations only occur if the first measurement is \emph{not} projective, $\phi_a \neq \pi/2$.

Notably, in Ref.~\cite{Dressel_StrengtheningWeakMeasurements_2018} we used precisely the same four-measurement circuit as in Fig.~\ref{fig:measurements} to obtain the real part of the OTOC $\text{Re}\la B(t)AB(t)A\ra$ itself. As such, once we add this fourth measurement to the circuit, we can use the previously derived four-measurement values $\xi^{\text{Re}BABA}_{a,b,a',b'} = 2\,\alpha_{\phi_{a},a}\,\alpha_{\phi_{b},b}\,\alpha_{\phi_{a'},a'}\,\alpha_{\phi_{b'},b'}-1$ as an alternative to the three-measurement values we introduced in Eq.~\eqref{eq:rebaba}. Similarly, as an alternative to Eq.~\eqref{eq:imbaba}, $\text{Im}\la BABA \ra$ can be obtained by averaging the four-measurement values $\xi^{\text{Im}BABA}_{a,b,a',b'} = 2\,\alpha_{\phi_{\tilde{a}},\tilde{a}}\,\alpha_{\phi_{b},b}\,\alpha_{\phi_{a'},a'}\,\alpha_{\phi_{b'},b'}$ over the circuit variation with a noninformative first measurement.

\section{Optimizing measurement strength}\label{sec:optimization}
All preceding derivations assumed arbitrary strength measurements and ideal probability distributions. However, in practice one measures realization frequencies in the lab, so both the experiment time and the statistical error must be taken into account. For a finite ensemble of realizations $N$ the squared deviation of the mean value, $(\Delta\bar{\xi})^2 = \sum_{k=1}^N (\xi_k-\bar{\xi})^2/N^2 \leq (\max_j \xi^2_j)/N$, is bounded from above by the largest averaged value. Here $k$ ranges over realizations and $j$ ranges over possible outcomes in one realization. Fixing the experiment time for one circuit realization and the admissible realization number $N$, we should minimize this deviation of the mean to conserve experimental resources.

As an example of this procedure, we examine the statistical error for one of the 16 QPD values:
\begin{align}\label{eq:qpdvalue}
&\text{Re}\la  \Pi_+^A \Pi_+^{B(t)} \Pi_+^A \Pi_+^{B(t)}\ra \xleftarrow{N\to \infty} \frac{1}{16N}\sum_{k=1}^N \big[ 3+3 \xi^A_k +3 \xi^B_k \nonumber \\
&\qquad + 4 \xi^{\text{Re}AB}_k + \xi^{BAB}_k + \xi^{ABA}_k + \xi^{\text{Re}BABA}_k \big],
\end{align}
where each $k$ is a particular realization of the measurement sequence $(a,b,a',b')$. To minimize the statistical error, we minimize the largest averaged value in this sum over all free parameters $\phi_a$, $\phi_b$, $\phi_{a'}$, and $\phi_{b'}$.
Numerical minimization yields different optimal strengths for each QPD value, with the one in Eq.~\eqref{eq:qpdvalue} having strengths
\begin{align}
\phi_{b} &= \phi_{a'} = \phi_{b'} = \pi/2, & \phi_a &\approx (0.67)\,\pi/2.
\end{align}
For all QPD values, all measurements are optimally projective except the first measurement, which has an optimum that is still reasonably strong ($\phi_a \approx (0.47)\,\pi/2$ or $\phi_a \approx (0.67) \pi/2$ ). A similar computation for the corresponding imaginary part $\text{Im}\la  \Pi_+^A \Pi_+^{B(t)} \Pi_+^A \Pi_+^{B(t)}\ra$ shows that projective measurements are always optimal for all measurements.


\section{Conclusions}\label{sec:conclusion}
For multi-qubit systems possessing local observables that square to the identity, we have reduced the problem of measuring the QPD behind the OTOC to that of determining eight independent real-valued correlators, in contrast to the $2^4$ complex parameters that ostensibly comprise the distribution. Six of these correlators can be constructed from one data set of the four-measurement circuit shown in Fig.~\ref{fig:measurements}. To minimize statistical error, all but the first measurement can be made projective, with only a slight strength reduction needed for the first measurement. The remaining two correlators can be obtained from a second data set from a slight variation of the same circuit that replaces the first measurement with a noninformative measurement and uses only three projective measurements. These simplifications greatly reduce the experimental difficulty for determining such a QPD.

The present work demonstrates that the same circuit used to sequentially measure a multi-qubit OTOC can also be used to determine all eight correlators needed to parametrize the QPD behind the OTOC. Thus, for qubits the QPD is no more difficult to measure with sequential measurements than the OTOC alone. We expect that measurements of this sort are presently attainable in modern quantum computing hardware. We also expect that aspects of this work may be extended to qutrits and higher-dimensional systems, where the assumption that observables square to the identity breaks down.

\begin{acknowledgments}
The authors are grateful for helpful discussions with Nicole Yunger Halpern and Mordecai Waegell. This work was partially supported by the
Army Research Office (ARO) grant No. W911NF-18-1-0178. JRGA was supported by a fellowship from the Grand Challenges Initiative at Chapman University, as well as kind hospitality from Franco Nori.
\end{acknowledgments}

\end{document}